\documentclass[superscriptaddress,prd,aps,
twocolumn,tightenlines,showpacs,amsmath,nofootinbib,floatfix]{revtex4}

\usepackage{graphicx}
\usepackage[varg]{txfonts}
\usepackage{array}
\usepackage{url}  
\def\n#1e#2n{#1\times10^{#2}}
\def\g#1{\gamma_{#1}}
\makeatletter
\def\slashchar#1{{\mathpalette\c@ncel{#1}}} 
\makeatother
\def\vsl{\slashchar{v}}
\def\vk{\mathbf{k}}
\def\vr{\mathbf{r}}

\begin{document}

\preprint{SHEP--0716}


\title{Semileptonic $bc$ to $cc$ Baryon Decay and Heavy Quark Spin Symmetry}

\author{Jonathan M.~Flynn}\affiliation{School of Physics and
  Astronomy, University of Southampton, Highfield, Southampton
  SO17~1BJ, UK}
\author{Juan~Nieves}\affiliation{Departamento de F\'isica At\'omica,
  Molecular y Nuclear, Universidad de Granada, E--18071 Granada,
  Spain}

\date{\today}

\begin{abstract}
We study the semileptonic decays of the lowest-lying $bc$ baryons to
the lowest-lying $cc$ baryons ($\Xi_{bc}^{(\prime*)}\to
\Xi_{cc}^{(*)}$ and $\Omega_{bc}^{(\prime*)}\to \Omega_{cc}^{(*)}$) ,
in the limit $m_b, m_c \gg \Lambda_\mathrm{QCD}$ and close to the zero
recoil point. The separate heavy quark spin symmetries make it
possible to describe all these decays using a single form factor. We
recover results derived previously by White and Savage in a manner
which we think is more straightforward and parallels the method
applied later to study $B_c$ semileptonic decays. We further discuss
the resemblance between the $bc$ baryon decays and those of $B_c$
mesons to $\eta_c$ and $J/\psi$ mesons and comment on the relation
between the slopes of the single functions describing each set of
decays. Our results can straightforwardly be applied to the decays of
$bb$ baryons to $bc$ baryons.
\end{abstract}

\maketitle

\section{Introduction}

The static theory for a system with two heavy quarks has infra-red
divergences which can be regulated by the kinetic energy term $\bar
h_Q (D^2/2 m_Q) h_Q$. This term breaks the heavy quark flavour
symmetry, but not the spin symmetry for each heavy quark flavour. The
spin symmetry is sufficient to derive relations between form factors
for decays of doubly heavy hadrons in the heavy quark limit, as was
first shown in~\cite{White:1991hz}. The consequences for semileptonic
decays of $B_c$ mesons were worked out in~\cite{Jenkins:1992nb}. Here
we extend the formalism to describe semileptonic decays of $bc$
baryons to $cc$ baryons. In Ref.~\cite{White:1991hz} the two heavy
quarks $Q$ in a $QQq$ baryon were treated as a point-like
colour-triplet anti-quark $\bar Q$ interacting with the light degrees
of freedom. We will compare our results with those obtained using this
diquark picture and make a link to the $B_c$ to $\eta_c$ and $J/\psi$
decays. For recent developments using the diquark picture
see~\cite{Mehen:2006vv,Hu:2005gf,Fleming:2005pd}.

We are interested in semileptonic decays of baryons containing two
heavy quarks and a light quark. Specifically we study the decays of
the cascade $bc$ baryons $\Xi_{bc}$, $\Xi'_{bc}$ and $\Xi^*_{bc}$ to
cascade $cc$ baryons $\Xi_{cc}$ and $\Xi^*_{cc}$. The quantum numbers
of these particles are listed in Table~\ref{tab:hh-baryons}. We find,
in agreement with~\cite{White:1991hz}, that in the heavy quark limit a
unique function describes the entire family of decays. This function
satisfies a normalisation condition (a consequence of vector current
conservation) at zero-recoil if the heavy quarks are degenerate. Our
results can be straightforwardly applied to the corrresponding decays
involving $\Omega$ baryons and also to the decays of $bb$ baryons to
$bc$ baryons. Some of these decays have also been studied in various
quark model
approaches~\cite{Sanchis-Lozano:1994vh,Guo:1998yj,Faessler:2001mr,Ebert:2004ck,Albertus:2006ya}.
\begin{table}[b]
\caption{Quantum numbers of double-heavy baryons. $S$ and $J^P$ are
the strangeness and the spin parity of the baryon, $I$ is the isospin
and $S_{hh'}^\pi$ is the spin parity of the heavy degrees of freedom,
well-defined in the infinite heavy mass limit. $l$ denotes a light $u$
or $d$ quark.}
\label{tab:hh-baryons}
\begin{ruledtabular}
\begin{tabular}[t]{>{$}c<{$}>{$}c<{$}>{$}c<{$}>{$}c<{$}>{$}c<{$}>{$}c<{$}>{\qquad$}c<{$}>{$}c<{$}>{$}c<{$}>{$}c<{$}>{$}c<{$}>{$}c<{$}}
& S & J^P & I & S_{hh}^\pi & & & S & J^P & I & S_{hh'}^\pi \\[0.5ex]
\hline
\Xi_{cc}      &  0 &\frac12^+ & \frac12 & 1^+& ccl &
 \Omega_{cc}   & -1 &\frac12^+ & 0 & 1^+ & ccs \\
\Xi^*_{cc}    &  0 &\frac32^+ & \frac12 & 1^+& ccl &
 \Omega^*_{cc} & -1 &\frac32^+ & 0 & 1^+ & ccs \\
\Xi'_{bc}     &  0 & \frac12^+ & \frac12 & 0^+& bcl &
 \Omega'_{bc}  & -1 & \frac12^+ & 0 & 0^+ & bcs \\
\Xi_{bc}      &  0 & \frac12^+ & \frac12 & 1^+& bcl &
 \Omega_{bc}   & -1 & \frac12^+ & 0 & 1^+ & bcs\\
\Xi^*_{bc}    &  0 & \frac32^+ & \frac12 & 1^+& bcl &
 \Omega^*_{bc} & -1 & \frac32^+ & 0 & 1^+ & bcs\\
\end{tabular}
\end{ruledtabular}
\end{table}

\section{Spin Symmetry}

The invariance of the effective Lagrangian under separate spin
rotations of the $b$ and $c$ quarks leads to relations between the
form factors for vector and axial-vector currents between the cascade
$bc$ baryons and cascade $cc$ baryons. These decays are induced by the
semileptonic weak decay of the $b$ quark to a $c$ quark. Near the zero
recoil point the velocities of the initial and final baryons are
approximately the same. If the momenta of the initial $bc$ and final
$cc$ baryons are $p_\mu = m_{bc} v_\mu$ and $p'_\mu = m_{cc}v'_\mu=
m_{cc}v_\mu + k_\mu$ respectively, then $k$ will be a small residual
momentum near the zero-recoil point. Since the final baryon is
on-shell, $k\cdot v = \mathcal{O}(1/m_{cc})$. We will work near
zero-recoil and thus neglect $v\cdot k$ below.

Heavy quark spin symmetry implies that all baryons with the same
flavour content listed in table~\ref{tab:hh-baryons} are degenerate.
The consequences of spin symmetry for weak matrix elements can be
derived using the ``trace formalism''~\cite{Falk:1990yz,MWbook}. To
represent the lowest-lying $S$-wave $bcq$ baryons we will use
wavefunctions comprising tensor products of Dirac matrices and
spinors, namely:
\begin{align}
\label{eq:Bprimebc}
B'_{bc} &=
 -\left[\frac{(1+\vsl)}2 \g5\right]_{\alpha\beta} u_\gamma(v,r)\\
\label{eq:Bbc}
B_{bc} &=
 \left[\frac{(1+\vsl)}2 \g\mu\right]_{\alpha\beta} \left[\frac1{\sqrt3}
 (v^\mu+\gamma^\mu)\g5  u(v,r)\right]_\gamma\\
\label{eq:Bstarbc}
B^*_{bc} = \Xi^*_{bc} &=
 \left[\frac{(1+\vsl)}2 \g\mu\right]_{\alpha\beta} u^\mu_\gamma(v,r)
\end{align}
where we have indicated Dirac indices $\alpha$, $\beta$ and $\gamma$
explicitly on the right-hand sides and $r$ is a helicity label for the
baryon. For the $B^*_{bc}$, $u^\mu_\gamma(v,r)$ is a Rarita-Schwinger
spinor. These wavefunctions can be considered as matrix elements of
the form $\langle0 | c_\alpha \bar{q^c}_\beta b_\gamma |
B^{(\prime*)}_{bc}\rangle$ where $\bar{q^c}=q^T C$ with $C$ the
charge-conjugation matrix. We couple the $c$ quark and light quark to
spin $0$ for the $B'_{bc}$ or $1$ for the $B_{bc}$ and $B^*_{bc}$
states. Under a Lorentz transformation, $\Lambda$, and $b$ and $c$
quark spin transformations $S_b$ and $S_c$, a wavefunction of the form
$\Gamma_{\alpha\beta}\, u_\gamma$ transforms as:
\begin{equation}
\label{eq:spintransfs}
\Gamma\,u \to S(\Lambda) \Gamma S^{-1}(\Lambda)\; S(\Lambda)u,
\quad
\Gamma\,u \to S_c \Gamma \, S_b u.
\end{equation}
The states in Eqs.~\eqref{eq:Bprimebc}, \eqref{eq:Bbc} and
\eqref{eq:Bstarbc} have a common normalisation $\bar u u
\mathrm{Tr}(\Gamma \overline\Gamma)$ and are mutually orthogonal.

To build states where the $b$ and $c$ quarks are coupled to definite
spin, we need the linear combinations
\begin{align}
\label{eq:Sbceq0}
|0;1/2,M\rangle_{bc} &= -\frac12 |0;1/2,M\rangle_{cq}
 + \frac{\sqrt3}2 |1;1/2,M\rangle_{cq}\\
\label{eq:Sbceq1}
|1;1/2,M\rangle_{bc} &= \frac{\sqrt3}2 |0;1/2,M\rangle_{cq}
 + \frac12 |1;1/2,M\rangle_{cq}\\
|1;3/2,M\rangle_{bc} &= |1;3/2,M\rangle_{cq}
\end{align}
where the second and third arguments are the total spin quantum
numbers of the baryon and the first argument denotes the total spin of
the $bc$ or $cq$ subsystem. We have chosen the relative phase of the
states in Eqs.~\eqref{eq:Sbceq0} and~\eqref{eq:Sbceq1} to agree with
that adopted above in Eqs.~\eqref{eq:Bprimebc} and~\eqref{eq:Bbc} (we
will comment again on this when constructing the $cc$ baryon states).
We have not used definite spin combinations for the $b$ and $c$ quarks
in Eqs.~\eqref{eq:Bprimebc} and \eqref{eq:Bbc}. This is to make both
the spin transformations on the heavy quarks and the Lorentz
transformation of the states convenient, making it straightforward to
build spin-invariant and Lorentz covariant quantities.

Finally we observe that we could have combined the $b$ quark with the
light quark to a definite spin in
Eqs.~\eqref{eq:Bprimebc}--\eqref{eq:Bstarbc}. This would clearly
interchange the spin transformations in Eq.~\eqref{eq:spintransfs}
(and alter the appearance of the matrix element expression in
Eq.~\eqref{eq:ME} below). Note also that when rewriting
Eq.~\eqref{eq:Sbceq0} with the roles of $b$ and $c$ exchanged, an
extra minus sign arises from the antisymmetry of the $S_{bc}=0$ state
under $b\leftrightarrow c$ interchange. Physical results should be
unaltered and we have checked that this is the case.

For the $cc$ baryons there are some differences because we have
two identical quarks. In this case the states are:
\begin{align}
\label{eq:Bprimecc}
B'_{cc} &=
 -\sqrt{\frac23}
 \left[\frac{(1+\vsl)}2 \g5\right]_{\alpha\beta} u_\gamma(v,r)\\
\label{eq:Bcc}
B_{cc} &=
 \sqrt2\left[\frac{(1+\vsl)}2 \g\mu\right]_{\alpha\beta}
 \left[\frac1{\sqrt3}(v^\mu+\gamma^\mu)\g5  u(v,r)\right]_\gamma\\
\label{eq:Bstarcc}
B^*_{cc} = \Xi^*_{cc} &=
 \sqrt{\frac12}
 \left[\frac{(1+\vsl)}2 \g\mu\right]_{\alpha\beta} u^\mu_\gamma(v,r)
\end{align}
Two comments are in order here. First, the two charm quarks can only
be in a symmetric spin-$1$ state and therefore $B'_{cc}$ and $B_{cc}$
correspond to the same baryon state $\Xi_{cc}$ (or $\Omega_{cc}$ if
the light quark is $s$). We can thus use either of them to build up
spin-invariants and we have confirmed that we obtain the same results
from each. Second, in the normalisation, there are two ways to
contract the charm quark indices, leading to $\bar u u
\mathrm{Tr}(\Gamma \overline\Gamma) + \bar u\, \Gamma\,
\overline\Gamma u$. In order to have the same normalisation as for the
$bc$ case, we have to include extra numerical factors as shown in
Eqs.~\eqref{eq:Bprimecc}--\eqref{eq:Bstarcc}. Note that the equality
between the $B'_{cc}$ and $B_{cc}$ states fixes the relative phase
between them.
 
We can now construct amplitudes for semileptonic cascade $bc$ to
cascade $cc$ baryon decays, determined by matrix elements of the weak
current $J^\mu = \bar c \gamma^\mu(1-\g5) b$. We first build
transition amplitudes between the $B^{(\prime*)}_{bc}$ and
$\Xi^{(*)}_{cc}$ states and subsequently take linear combinations to
obtain transitions from $\Xi^{(\prime*)}_{bc}$ states. The most
general form for the matrix element respecting the heavy quark spin
symmetry is\footnote{If the roles of the $b$ and $c$ quarks were
interchanged, the matrix element would read $\bar u_{cc}u_{bc}
\mathrm{Tr}[\Gamma_{bc}\Omega \overline\Gamma_{cc}\gamma^\mu(1-\g5)] +
\bar u_{cc}\gamma^\mu(1-\g5)
\Gamma_{bc}\Omega\overline\Gamma_{cc}u_{bc}$.}
\begin{multline}
\label{eq:ME}
\langle\Xi^{(*)}_{cc},v,k,M'|J^\mu(0)|B^{(\prime*)}_{bc},v,M\rangle \\
\begin{aligned}
 &=\bar u_{cc}(v,k,M') \gamma^\mu(1-\g5) u_{bc}(v,M)
   \mathrm{Tr}[\Gamma_{bc}\Omega \overline\Gamma_{cc}]\\ 
 & \quad + \bar u_{cc}(v,k,M')
      \Gamma_{bc}\Omega\overline\Gamma_{cc} \gamma^\mu(1-\g5) u_{bc}(v,M)
\end{aligned}
\end{multline}
where $M$ and $M'$ are the helicities of the initial and final states
and $\Omega = -\eta(\omega)/\sqrt2$, with $\omega=v\cdot v'$. We use
the standard relativistic normalisation for hadronic states and our
spinors satisfy $\bar u u = 2 m$, $\bar u^\mu u_\mu = - 2 m$ where $m$
is the mass of the state. Terms with a factor of $\vsl$ can be omitted
because of the equations of motion ($\vsl u=u$, $\vsl\Gamma=\Gamma$,
$\g\mu u^\mu=0$, $v_\mu u^\mu=0$), while terms with $\slashchar k$
will always lead to contributions proportional to $v\cdot k$ which is
set to $0$ at the order we are working. We also make use of the
relations $\bar u \g\mu u = \bar u v_\mu u$, $\bar u \g5 u = 0$, $\bar
u \slashchar k u =0$ and $\bar u \slashchar k \g\mu \g5 u = - \bar u
\slashchar k v_\mu \g5 u$. Our results for cascade $bc$ to cascade
$cc$ transition matrix elements are:
\begin{align}
\label{eq:XibctoXicc}
\Xi_{bc}\to \Xi_{cc} \qquad &
 \eta\, \bar u_{cc}\left(2\gamma^\mu -\frac43\gamma^\mu\g5\right) u_{bc}\\
\Xi'_{bc}\to \Xi_{cc} \qquad &
 \frac{-2}{\sqrt3} \eta\, \bar u_{cc}\left(-\gamma^\mu\g5\right) u_{bc}\\
\Xi_{bc}\to \Xi^*_{cc} \qquad &
 \frac{-2}{\sqrt3} \eta\, \bar u^\mu_{cc} u_{bc}\\
\Xi'_{bc}\to \Xi^*_{cc} \qquad &
 -2 \eta\, \bar u^\mu_{cc} u_{bc}\\
\label{eq:XibcstartoXicc}
\Xi^*_{bc}\to \Xi_{cc} \qquad &
 \frac{-2}{\sqrt3} \eta\, \bar u_{cc} u^\mu_{bc}\\
\Xi^*_{bc}\to \Xi^*_{cc} \qquad &
 -2 \eta\, \bar u^\lambda_{cc}\left(\gamma^\mu-\gamma^\mu\g5\right)
 u_{bc\,\lambda}
\end{align}
If the $b$ and $c$ quarks become degenerate, then vector current
conservation ensures that $\eta(1)=1$.

The consequences of taking the heavy quark limit for semileptonic
decays of baryons with two heavy quarks were considered some time ago
by Savage and White~\cite{White:1991hz}. They adopted an approach
where the two heavy quarks bind into a colour antitriplet which
appears as a pointlike colour source to the light degrees of freedom.
Applying the ``superflavor'' formalism of Georgi and
Wise~\cite{Georgi:1990ak,Savage:1990di,Carone:1990pv} allowed the
matrix elements of the heavy-flavour-changing weak current to be
evaluated between different baryon states. We find two differences to
their results which cannot be eliminated by redefining the phases of
the physical states. One difference, already pointed out
in~\cite{Sanchis-Lozano:1993kh}, is for the spin-$3/2$ to spin-$1/2$
transition in Eq.~\eqref{eq:XibcstartoXicc}, where they find a
vanishing weak transition matrix element, while ours is non-zero. The
second difference is the relative sign of the vector and axial
contributions in the $\Xi_{bc}\to\Xi_{cc}$ transition of
Eq.~\eqref{eq:XibctoXicc}. This does not affect the differential decay
rate although it could change angular correlations between the
outgoing charged lepton and baryon.

Spin symmetry for both the $b$ and $c$ quarks enormously simplifies
the description of all of the above transitions in the heavy quark
limit and near the zero recoil point. All the weak transition matrix
elements are given in terms of a single universal function. Lorentz
covariance alone allows a large number of form factors (six form
factors to describe $\Xi_{bc}\to\Xi_{cc}$, another six for
$\Xi'_{bc}\to\Xi_{cc}$, eight each for $\Xi_{bc}\to\Xi^*_{cc}$,
$\Xi'_{bc}\to\Xi^*_{cc}$ and $\Xi^*_{bc}\to\Xi_{cc}$, and even more
for $\Xi^*_{bc}\to\Xi^*_{cc}$). The spin symmetry provides further
simplifications beyond those coming from working at $v'=v$. For
example, the transitions $\Xi^{(\prime)}_{bc}\to \Xi_{cc}$ are
each described by six form factors in general, corresponding to the
structures $v^\mu-\gamma^\mu$, $v'^\mu-\gamma^\mu$, $\gamma^\mu$,
$v^\mu\g5$, $v'^\mu\g5$ and $\gamma^\mu\g5$. At the zero recoil point
only $\gamma^\mu$ and $\gamma^\mu\g5$ survive, leaving four form
factors to describe these two decays. Spin symmetry reduces this to a
single function $\eta$, which also describes the rest of the
transitions shown above.

\section{Diquark Picture and Link to $B_c$ Meson Decays}

Up to now we have used only the separate spin symmetries for the heavy
charm and bottom quarks and our results are completely
model-independent. Now we will use constituent quark model ideas to
estimate the scale of variation of the form factors and to make a link
to $B_c$ to $\eta_c$ and $J/\psi$ semileptonic decays.

The form factor $\eta$ is calculable in terms of the overlap of the
spatial wave functions of the $bcq$ and $ccq$ baryon states.
Considering the $\Xi_{bc}\to\Xi_{cc}$ transition with the initial
baryon at rest, we can find $\eta$
using
\begin{multline}
\eta(\omega) = \int d^3 r_1 d^3 r_2 \exp[-i\vk\cdot\vr_{12}/2] \\
 \times [\Psi^\Xi_{cc}(r_1,r_2,r_{12})]^*\, \Psi^\Xi_{bc}(r_1,r_2,r_{12})
\end{multline}
where $r_{1,2}$ are the distances between each of the heavy quarks and
the light quark, while $r_{12}$ is the heavy quark separation. The
wave functions depend on distances because we are assuming that the
lowest-lying baryons are purely $S$-wave and so the integral depends
on $\vk^2 = m_{\Xi_{cc}}^2[\omega^2-1]$ (see Eq.~(34)
in~\cite{Albertus:2006ya}).

If the distance between the two heavy quarks is much smaller than the
distance of the light quark from either heavy quark, as expected in
the heavy mass limit of a strong Coulomb binding potential where the
radius of the $QQ$ bound state should decrease as $1/m_Q$, then the
baryon wave functions can be approximated by (see appendix~B
of~\cite{Albertus:2006ya})
\begin{equation}
\Psi^\Xi_{Qc}(r_1,r_2,r_{12}) =  \Phi_{Qc}(r_{12}) \phi(r_Q)
\end{equation}
where $r_Q$ is the distance of the light quark from the centre of mass
of the two heavy quarks. We ignore all spin-dependent interactions
which are suppressed by inverse powers of heavy quark masses, allowing
us to drop the superscript $\Xi$ from now on, and making all
interquark potentials flavour independent. $\Phi_{Qc}$ is the
ground-state wavefunction of the $Qc$ diquark, while $\phi$ is the
ground-state wavefunction for the relative motion of the light quark
and a pointlike diquark of infinite mass with a potential which is
twice the quark-quark potential. In these circumstances we have
\begin{multline}
\eta(\omega) = \int d^3 r_{12} \exp[-i\vk\cdot\vr_{12}/2]
  \Phi_{cc}^*(r_{12}) \Phi_{bc}(r_{12})\\
 \times \int d^3 r \phi^*(r)\phi(r)
\end{multline}
where $\vr = \vr_c$ and in the $d^3 r$ integral we have replaced
$\phi(r_b)$ by $\phi(r)$ since $\vr_b = \vr_c + \mathcal{O}(\vr_{12})$.
This approximation leads to uncertainties of $\mathcal{O}(r_{12}^2)$
after integration. The $d^3 r$ integration then gives $1$ and thus
\begin{equation}
\label{eq:eta}
\eta(\omega) = \int d^3 r_{12} \exp[-i\vk\cdot\vr_{12}/2]
  \Phi_{cc}^*(r_{12}) \Phi_{bc}(r_{12})
\end{equation}
which has an identical form to Eq.~(4.11)
in~\cite{Jenkins:1992nb}, where the unique form factor $\Delta$
describing the $B_c$ to $\eta_c$ and $J/\psi$ semileptonic decays is
given in terms of wavefunctions of the $\bar b c$ and $\bar c c$ bound
states\footnote{We believe that there should not be an explicit factor
of $2$ in~(4.11) of~\cite{Jenkins:1992nb}. This factor does not appear
in the corresponding expressions
in~\cite{Gershtein:1994jw,Sanchis-Lozano:1994vh,Choi:1998rs,Colangelo:1999zn,Ivanov:2000aj,Hernandez:2006gt}.}.
This does not mean that $\eta$ and $\Delta$ are identical because the
$QQ$ and $Q\bar Q$ potentials used to compute the diquark and meson
wavefunctions are not the same. For example a $\lambda_i \lambda_j$
colour dependence ($\lambda_i$ are the usual Gell-Mann matrices) would
lead to $V^{QQ}=V^{Q\bar Q}/2$.

Assuming Coulomb wavefunctions, $\Phi_{Qc}(r)\propto e^{-r/a_Q}$, with
the diquark radius $a_Q \propto 1/(\beta \mu_Q)$, where $\mu_Q$ is the
$Qc$ reduced mass and $\beta$ is the strength of the $-1/r$ potential,
we find
\begin{equation}
\eta(\omega) = 8 \frac{a_b^{3/2}a_c^{3/2}}{(a_b+a_c)^3}
 \left[1+\frac{\vk^2 a_b^2 a_c^2}{4(a_b+a_c)^2}\right]^{-2}
\end{equation}
which agrees with the expression given in Eq.~(3)
of~\cite{White:1991hz} and clearly resembles Eq.~(4.12)
of~\cite{Jenkins:1992nb}. Assuming $V^{QQ}=V^{Q\bar Q}/2$, we would
expect the $B_c$ and $\eta_c$ radii $a_0$ and $a_\eta$ introduced
in~\cite{Jenkins:1992nb} to be approximately one half of $a_b$ and
$a_c$ respectively. The $\omega^2$ slopes of the form factors $\Delta$
and $\eta$ would then be in the ratio $1$ to $4
(m_{\Xi_{cc}}/m_{\eta_c})^2 \sim 6$.

\begin{figure}
\includegraphics[width=\hsize]{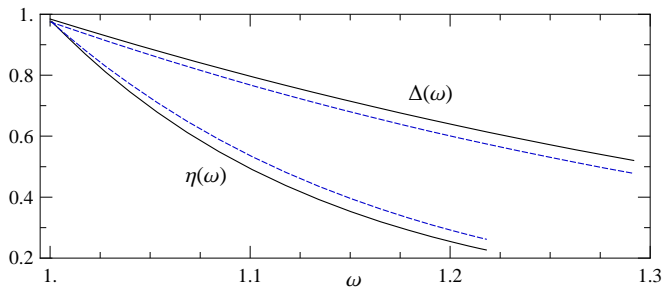}
\caption{Form factors in the heavy quark limit: $\eta(\omega)$ for
cascade $bc$ to cascade $cc$ baryon decays and $\Delta(\omega)$ for
$B_c\to\eta_c, J/\psi$ decays, calculated from a nonrelativistic quark
model~\cite{Hernandez:2006gt,Albertus:2006ya} (using the AL1
potential). The solid lines are calculated from the wavefunction
overlaps, illustrated for $\eta(\omega)$ in Eq.~\eqref{eq:eta}, while
the dashed lines are constructed from appropriate combinations of form
factors: for $\eta$ we consider $(F_1+F_2+F_3)/2$, where $F_{1,2,3}$
are defined in Eq.~(23) of~\cite{Albertus:2006ya}, while for $\Delta$
we use $\Sigma_1^{(0)}$ defined in Eq.~(52)
of~\cite{Hernandez:2006gt}. The solid and dashed curves should agree
close to zero recoil ($\omega \to 1$).}
\label{fig:eta-and-delta}
\end{figure}

To check the use of Coulomb wavefunctions and the slope prediction, we
have calculated $\eta$ and $\Delta$ using wavefunctions from a
nonrelativistic quark model~\cite{Hernandez:2006gt,Albertus:2006ya}
and show the results in Fig.~\ref{fig:eta-and-delta}. The $\omega^2$
slope of the $\Delta$ form factor is indeed smaller than that of
$\eta$, but the ratio is around $1$ to $3$ rather than $1$ to $6$, so
there are significant corrections to the Coulomb wavefunction
description.

\section{Conclusion}

We have studied the semileptonic decays of the lowest-lying $bc$
baryons to lowest-lying $cc$ baryons in the limit $m_b, m_c \gg
\Lambda_\mathrm{QCD}$ and close to the zero recoil point. The separate
heavy quark spin symmetries make it possible to describe all these
decays using a single form factor. We have discussed the resemblance
of the $bc$ baryon decays to those of $B_c$ mesons to $\eta_c$
and $J/\psi$ mesons and commented on the relation between the slopes
of the single functions describing each set of decays. Lattice QCD
simulations work best near the zero-recoil point and thus are
well-suited to check the validity of the results.

We studied specifically the semileptonic decays of cascade $bc$
baryons to cascade $cc$ baryons. Our results can be straightforwardly
applied also to the corrresponding decays involving $\Omega$ baryons
as well as to the decays of $bb$ baryons to $bc$ baryons. It is also
straightforward to extend the analysis to transitions involving the
heavy-to-light weak current, using the $bc$ baryon wavefunctions
defined in Eqs.~\eqref{eq:Bprimebc}, \eqref{eq:Bbc} and
\eqref{eq:Bstarbc} together with the usual spinor wavefunction for a
single heavy quark baryon. For example, to study
$\Xi^{(\prime*)}_{bc}\to \Lambda_b$ semileptonic decays, we would
evaluate expressions like $\bar u_b u_{bc}\,
\mathrm{Tr}[\gamma^\mu(1-\g5)\Gamma_{bc}\Omega]$ where $\Omega =
\Omega_1 + \slashchar k \Omega_2$ and $u_b$ is the spinor for the
$\Lambda_b$.

\section*{Acknowledgements}

We thank E.~Hernandez and J-M.~ Verde-Velasco for providing quark
model wave functions and form factors. JMF thanks the Departamento de
F\'isica At\'omica, Molecular y Nuclear, Universidad de Granada for
hospitality. We acknowledge grants MEC FIS2005-00810, MEC
SAB2005-0163, Junta de Andalucia FQM0225, PPARC PP/D000211/1 and EU
FLAVIAnet MRTN-CT-2006-035482.

\bibliographystyle{apsrev}
\bibliography{hh-hadron}

\end{document}